# Geographical veracity of indicators derived from mobile phone data.

Maarten Vanhoof
Open Lab, Newcastle University, UK
Orange Labs, Paris, FR
M.vanhoof1@newcastle.ac.uk

Thomas Plötz
Open Lab, Newcastle University, UK
School of interactive computing,
Georgia Institute of Technology, USA

Zbigniew Smoreda
Orange Labs, Paris, FR

**ABSTRACT**

*In this contribution we summarize insights on the geographical veracity of using mobile phone data to create (statistical) indicators. We focus on problems that persist with spatial allocation, spatial delineation and spatial aggregation of information obtained from mobile phone data. For each of the cases, we offer insights from our works on a French CDR dataset and propose both short and long term solutions. As such, we aim at offering a list of challenges, and a roadmap for future work on the topic.*

## 1. Introduction

Ever since the first analyses on mobile phone data, individual indicators have been constructed to describe mobility patterns [1], presence patterns [2], aspects of social networks [3] or even personal characteristics [4]. Although originally used to explore the characteristics of large-scale datasets, several applications of mobile phone indicators have been developed over time. And this to a degree that real-time monitoring of socio-economic landscapes based on mobile phone indicators seems now plausible [5], [6].

One logical application of mobile phone indicators is to complement national statistics. As they capture behavior for large populations, the combination with census data (which typically offer contextual information) opens up exciting research opportunities. [6], for example, investigate the relation between deprivation of the neighborhood and the diversity of individual mobility by pairing national statistics with mobile phone data of 20 million users in France. Apart from the obvious confrontation with official statistics, other applications integrate mobile phone indicators for a multitude of purposes. Mobility indicators, for instance, are used to segment users in epidemiological model [7] or to express their 'vulnerability' to wrongful home detection [8]. Indicators on calling behavior can be used to determine social influencers, to segment customers for marketing purposes [9], to annotate social networks [10], etc..

Here, our argument is that, despite many applications, important questions remain on how to evaluate the veracity of indicators produced from mobile phone data. Specifically, we focus on three different spatial aspects that influence uncertainty and error when creating or using mobile phone based indicators: spatial allocation, spatial delineation and special aggregation.

## 2. Spatial allocation

Often, individual mobile phone users have to be allocated in space in order to enable further analysis of indicators derived from their mobile phone use. Typically, such allocation happens by means of home detection algorithms that incur the most plausible cell-tower to cover a user's home location based on rather simple heuristics (e.g. maximum activities during nighttime). Until now, little work has been done to assess and compare the performance of different heuristics as good validation datasets are hard to find (a.o. because of large population size, differences in cover grids with census, etc.). As a consequence, it is difficult to evaluate the uncertainty for allocating users, and thus indicators, to geographical areas [8].

To address the spatial allocation problem, in [8] we investigated the performance of five simple heuristics for home detection on a French CDR dataset. We applied each heuristic to almost 120 million mobility traces derived from one month of mobile phone data and compared results between them. Additionally, we collaborated with the French National Statistics Office (INSEE) to create a validation dataset describing the French population at the spatial resolution on the cell-tower grid provided by the operator. This allowed us to compare full population numbers with estimated population based on mobile phone data and thus assess performance of the different algorithms both in time as in space (figure 1).

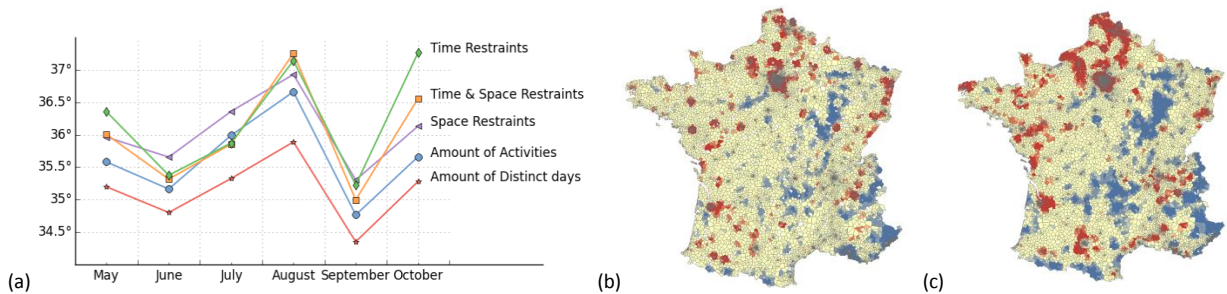

Figure 1: (a) Cosinus Similarity Values (in °) for the comparison between the vectors of (i) the number of detected user homes by the given home detection algorithm and (ii) the total population in the validation dataset based on census data. Vectors are made up of the totals of (i) and (ii) for each cell-tower. A value of 0° denotes the highest possible similarity between both vectors, 90° indicates the lowest similarity in orientation whereas 180° degrees refers to an opposite orientation. (b) and (c) Hotspots (red) and coldspots (blue) defined by the 90+% interval of the Getis-Ord Gi* statistic for (a) the number of detected homes by the amount of activities algorithm in June and (c) the population numbers of the validation dataset based on census data. The map is a made up by the Voronoi tesselation based on cell-tower locations.

## 3. Spatial delineation

Uneven delineations of space are a second problem when pursuing analysis with mobile phone indicators. Cell-tower cover areas, for example, typically have different boundaries compared to administrative regions, resulting in translation and comparisons problems. Uneven spatial delineations also exist between cell-tower cover areas. Cell-tower density in high population density areas, for instance, is typically higher, resulting in smaller cover areas compared to lower population density areas. Logically, this is directly relevant for population density estimations [11], the creation of mobility indicators [12], or parameter estimation in statistical analysis (e.g. urban scaling laws [13]).

In [12] we address the spatial delineation problem by assessing its effect on the creation of one mobility indicator from mobile phone data. As illustrated in figure 3 (a), we unveiled the calculation of mobility entropy (as proposed in [1], [6]) to be dependent on the density of cell-towers and thus the spatial delineation of cell-tower cover areas. To counter this, we propose a correction of the mobility entropy indicator. Application to a French CDR dataset learns that our proposed Corrected Mobility Entropy is less dependent on cell-tower densities and exhibits a different spatial pattern. We find suburban areas of large cities to depict the highest diversity of movement, compared to city centers for the standard mobility entropy (figure 3 (b) and (c)), as well as a clear decrease in mobility entropy with city size.

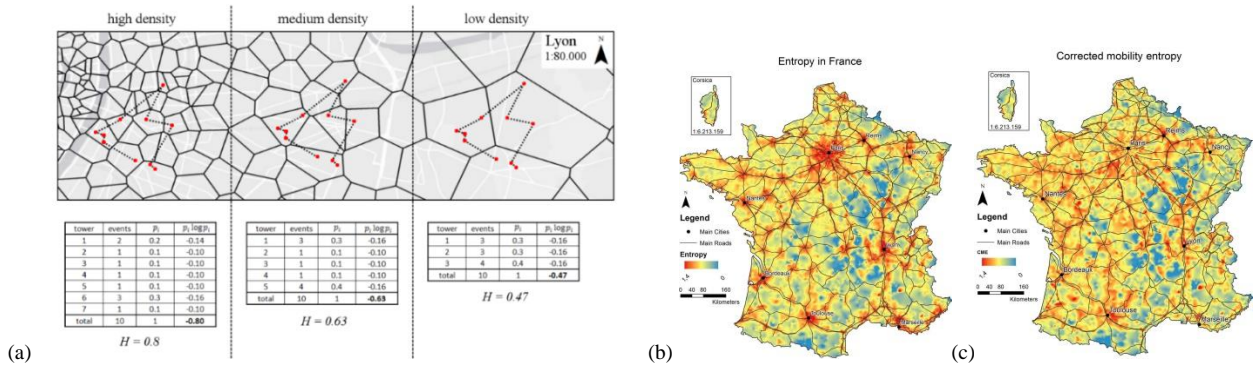

Figure 3: (a) Illustration of the effect of spatial delineation on the calculation of the temporally-uncorrelated entropy (H). The mobility entropy for the same path (dotted line) of a user in three different density settings is calculated and shown to be different. Spatial patterns of mobility entropy values (a) and corrected mobility entropy values (b). Values are calculated for all cell-towers in France as the average value of all users that have a 'home' at these cell-towers.

## 4. Spatial aggregation

A third problem consists in the spatial aggregation of mobile phone indicators. Although related to the spatial delineation problem, with spatial aggregation we do not mean by which spatial delineation one aggregates. Rather we wonder how the properties of indicators change when aggregating to another spatial level. This is especially relevant for mobile phone based indicators as the choice of scale will imply unintended selective filtering (e.g. highly active persons, communities), a change in representativeness of population (e.g. unequal market shares for single operators between city and rural areas), a change in human behavior (e.g. short-distance vs. long-distance trips) and infrastructural context (e.g. transport infrastructure); all of which are typically unaddressed in studies using mobile phone data.

Currently, there is little evidence that research is taking into account any (plausible) effects of spatial aggregation. Even though, in our work, we find clear indications that different scales imply different findings like, for example, in the relations between mobile phone indicators, home detection algorithms and census information (table 1). It remains an open question as to what mechanisms are responsible.

Table 1: Correlation coefficients between different indicators at different spatial levels in France. Administrative level is the official IRIS level and municipality level is the commune in French official statistics. Cell-tower level corresponds to estimated cover areas of Orange cell-towers in France. EDI is the European Deprivation Index, Mobility Entropy is as calculated in [12]. Distinct days home detection is also used in figure 1 and described in [8].

| Correlation coefficient for | Cell-tower Level | Administrative Level | Municipality Level | Scale difference |
|---|---|---|---|---|
| **Mobility Entropy vs. EDI** |  | -0.03 | -0.43 | -0.40 |
| **Distinct Days Home Detection vs. Population in Census data** | 0.62 | 0.92 |  | +0.30 |

## 5. Roadmap

Ultimately, we recognize that procedures to create mobile phone based indicators are being institutionalized in, at least, two ways. On the one hand, there is published research and the development of open source software packages like, for example, the python toolbox bandicoot [14]. Such packages allow researchers to easily calculate indicators but their recurrent use might eventually lead to ill-considered applications being legitimized by the toolbox instead of the specific nature of the research case. In this perspective, it's a bit uncanny that current open-source packages offer little consideration to the assessment of uncertainty or error and, as a consequence, to potentially wrongful deployment or interpretation.

On the other hand, national statistic offices are putting considerable efforts to investigate the usability of mobile phone indicators within national statistics. Here, extensive validation of mobile phone indicators is the norm, given that, when ensured by the office, they will directly inform policy decisions. The problem, however, is that the requirements, procedures and measures used to validate and publish official statistics are not (yet) adapted to the nature of current big data sets, including mobile phone data. The assessment of veracity of mobile phone data hence becomes a matter of re-invention of the national statistics services, which likely will take several years before being fully operational.

Within this context, we believe it is necessary for future work to investigate, or at least openly communicate on, the different forms of uncertainties and errors that occur because of spatial allocation, spatial delineation, and/or spatial aggregation. Table 2 offers a more specific roadmap for actions that can be taken in the short and long term to prevent naïve application of mobile phone based indicators.

Table 2: Proposed short term actions and long term solutions to the spatial allocation, delineation and aggregation problem.

| Problem | Short term actions | Long term solutions |
|---|---|---|
| **Spatial allocation** | - Investigate errors that come with spatial allocation<br>- Test heuristics for home detection on different databases<br>- Design surveys to gather ground truth at individual level | - Understand how characteristics of current mobile phone use (frequency, social context, etc.) influence spatial allocation<br>- Standardize error assessment for spatial allocation |
| **Spatial delineation** | - Assess the influence of spatial delineation on indicators<br>- Develop techniques that allow translation between different spatial delineations | - Develop standard practices that incorporate the effect of spatial delineation on indicators<br>- Reflect on possibilities to standardize spatial delineations like cell-tower areas, administrative boundaries, etc.) |
| **Spatial aggregation** | - Develop techniques that define optimal spatial scale for studying a specific process (be it theoretical or empirical)<br>- Develop techniques that express sensitivity of data (interpretation) to spatial aggregation | - Develop techniques that are capable of dealing with changing nature of observations when spatially aggregating<br>- Understand how the problem of spatial aggregation changes in time due to changing mobile phone use and human behavior |

**References**

[1] C. Song, Z. Qu, N. Blumm, and A.-L. Barabási, "Limits of Predictability in Human Mobility," *Science (80-. ).*, vol. 327, no. 5968, pp. 1018–1021, Feb. 2010.

[2] M. Janzen, M. Vanhoof, and K. W. Axhausen, "Estimating Long-Distance Travel Demand with Mobile Phone Billing Data," *16th Swiss Transp.*, 2016.

[3] D. Wang, D. Pedreschi, C. Song, F. Giannotti, and A.-L. Barabasi, "Human mobility, social ties, and link prediction," in *Proceedings of the 17th ACM SIGKDD international conference on Knowledge discovery and data mining*, 2011, pp. 1100–1108.

[4] Y.-A. de Montjoye, J. Quoidbach, F. Robic, and A. S. Pentland, "Predicting personality using novel mobile phone-based metrics," in *International Conference on Social Computing, Behavioral-Cultural Modeling, and Prediction*, 2013, pp. 48–55.

[5] F. Giannotti, D. Pedreschi, A. Pentland, P. Lukowicz, D. Kossmann, J. Crowley, and D. Helbing, "A planetary nervous system for social mining and collective awareness," *Eur. Phys. J. Spec. Top.*, vol. 214, no. 1, pp. 49–75, 2012.

[6] L. Pappalardo, M. Vanhoof, L. Gabrielli, Z. Smoreda, D. Pedreschi, and F. Giannotti, "An analytical framework to nowcast well-being using mobile phone data," *Int. J. Data Sci. Anal.*, Jun. 2016.

[7] A. Lima, V. Pejovic, L. Rossi, M. Musolesi, and M. Gonzalez, "Progmosis: Evaluating Risky Individual Behavior During Epidemics Using Mobile Network Data," *arXiv Prepr. arXiv*, vol. abs/1504.0, 2015.

[8] M. Vanhoof, F. Reis, Z. Smoreda, and T. Plötz, "Investigating Performance and Spatial Uncertainty of Home Detection Criteria for CDR data."

[9] P. Sundsøy, J. Bjelland, A. M. Iqbal, Y.-A. de Montjoye, and others, "Big Data-Driven Marketing: How machine learning outperforms marketers' gut-feeling," in *International Conference on Social Computing, Behavioral-Cultural Modeling, and Prediction*, 2014, pp. 367–374.

[10] J. L. Toole, C. Herrera-Yaqüe, C. M. Schneider, and M. C. González, "Coupling human mobility and social ties," *J. R. Soc. Interface*, vol. 12, no. 105, 2015.

[11] P. Deville, C. Linard, S. Martin, M. Gilbert, F. R. Stevens, A. E. Gaughan, V. D. Blondel, and A. J. Tatem, "Dynamic population mapping using mobile phone data," *Proc. Natl. Acad. Sci.*, vol. 111, no. 45, pp. 15888–15893, Nov. 2014.

[12] M. Vanhoof, W. Schoors, A. Van Rompaey, T. Plötz, and Z. Smoreda, "Correcting Mobility Entropy for Large-Scale Comparison of Individual Movement Patterns."

[13] C. Cottineau, E. Hatna, E. Arcaute, and M. Batty, "Paradoxical interpretations of urban scaling laws," *arXiv Prepr. arXiv1507.07878*, 2015.

[14] Y.-A. de Montjoye, L. Rocher, and A. S. Pentland, "bandicoot: a Python Toolbox for Mobile Phone Metadata," *J. Mach. Learn. Res.*, vol. 17, no. 175, pp. 1–5, 2016.